\documentclass[reqno]{amsart}
\usepackage{amsmath,amssymb}
 \theoremstyle{plain}

\theoremstyle{definition}

\theoremstyle{remark}

\numberwithin{theorem}{section} \numberwithin{equation}{section}

\begin{document}

\vspace*{1cm}
\begin{center}
{\Large \bf  A remark on rational isochronous potentials}
\\*[4ex]
{\bf O.A. Chalykh$^{\dagger}$ and A.P. Veselov$^{\ddagger,\star}$}
\\*[2ex]
\end{center}

\noindent $^\dagger$ Department of Applied Mathematics, University
of Leeds, Leeds, LS2 9JT, UK

\noindent $^\ddagger$ Department of Mathematical Sciences,
Loughborough University,\\ Loughborough,  LE11 3TU, UK

\noindent $^\star$ Landau Institute for Theoretical Physics,
Kosygina 2, Moscow, 117940, Russia

\noindent E-mail addresses: o.chalykh@mail.ru,
A.P.Veselov@lboro.ac.uk

\vspace*{0.5cm}

\centerline{\it Dedicated to Francesco Calogero on his
$70^{\text{th}}$ birthday}

\vspace*{0.5cm}

{\small  {\bf Abstract.} We consider the rational potentials of
the one-dimensional mechanical systems, which have a family of
periodic solutions with the same period (isochronous potentials). We
prove that up to a shift and adding a constant all such potentials
have the form $U(x) = \frac{1}{2}  \omega^2 x^2$ or  $U(x) = \frac{1}{8}
\omega^2 x ^2 + c^2 x^{-2}.$ }

\section{Introduction}

Consider one-dimensional mechanical system with the Hamiltonian
$$H = \frac{1}{2} p^2 + U(q)$$ and ask when such a system has a family of
oscillatory solutions with the {\it same } period $T$. In that
case we will call the corresponding potential {\it isochronous}.
This is a very old question going back to Huygens  and discussed in the classical
textbooks on mechanics (see  Appell \cite{A}, section 213 and Landau-Lifschitz \cite{LL}, section 12) 
and in several papers of which we mention \cite{U}, \cite{CMV}.

The aim of this note is to point out that although the space of
the isochronous potentials is pretty large (see the general
description in the next section) only few of them are rational
functions. Our main result is the following

{\bf Theorem.} {\it Up to a shift $x \rightarrow x+a$ and adding a
constant all rational isochronous potentials have the form $U(x) = \frac{1}{2} \omega^2 x^2$ or
\begin{equation}
\label{U} U(x) = \frac{1}{8} \omega^2 x ^2 + \frac{c^2} {x^2},
\end{equation}
where $c$ is any non-zero constant and $\omega = 2 \pi /T$. }

We had not found this result in the literature although some
considerations of \cite{CMV} are very close to it. The proof is
actually not difficult anyway (see next section). The reason we
find this result important is that it demonstrates an exceptional
nature of the Calogero-Moser system \cite {C}.  This system
describes the pairwise interaction of the particles on the line
with the potential $$V(x_1, \dots, x_n) = \frac{1}{2}  \omega^2 \sum x_i ^2  +
C \sum (x_i - x_j)^{-2}.$$ It is known \cite{OP} that the motion
of this system is periodic with the same period $T=\pi/\omega.$
In two-particle case this reduces to the isochronicity of the
potential (\ref{U}).

The isochronicity of the family (\ref{U}) can be proved
in many ways, of which the most simple one is probably the following. Consider
two-dimensional isotropic oscillator with the Hamiltonian $$H =
\frac{1}{2} (p_1^2 + p_2^2) +  \frac{1}{2} \omega^2 (q_1^2 + q_2^2).$$ The dynamics of the
distance $r = (q_1^2 + q_2^2)^{1/2}$ is described by the
one-dimensional system with effective potential $U(r) = \frac{1}{2} \omega^2 r
^2 + C r^{-2},$ where constant $C$ is half of the square of the
angular momentum. Now since the motion of the isotropic harmonic
oscillator is periodic with the period $T = 2\pi/\omega$ and the
orbits are ellipses the motion in the potential $U(r)$ is also
periodic with the period $T= \pi /\omega$ (coefficient 1/2 is due
to the central symmetry of the ellipses).

We do not know much about multidimensional generalisations of our
result. Some interesting examples can be found in the recent paper
\cite{C2} by Calogero, who did a lot to revitalise this area. Our
interest to this problem was to a large extent stimulated by his
enthusiasm.

\section{Proof of the Theorem}

Let us first note that isochronous motion is possible only in the interval where
the potential $U(x)$ has the only (and thus global) minimum. Indeed the presence of any other extrema
will lead to the divergence of the period given by the integral 
$$T = \sqrt{2} \int \frac{dx}{\sqrt{E - U(x)}}$$
(see \cite{LL}), which must be independent on the energy $E$ (by analyticity arguments).

Let us assume for convenience that the period of oscillations $T =
\sqrt{2} \pi$ (so that $\omega = \sqrt{2}$ and the corresponding harmonic potential is simply $U=x^2$) and that potential energy of the equilibrium is zero. This
means that our rational function $y = U(x)$ has a local minimum $x
= x_0$ with $y_0 = U(x_0) = 0.$

Let $x_1(y)$ and $x_2(y)$  be corresponding local solutions of the
equation $U(x) = y$, i.e. the branches of the inverse function
$U^{-1}(y)$ bifurcating from $x_0$: $x_1(0) = x_2(0) = x_0, x_1
(y) \le x_2(y)$ for positive $y$.

The classical criterium of isochronicity is that the difference
$$\Delta(y) = x_2(y) - x_1(y)$$ is exactly the same as for the
corresponding harmonic oscillator (see \cite{LL}, section 12). In particular
for our choice of $T$ we must have
\begin{equation}
\label{C} x_2(y) - x_1(y) = 2 \sqrt y
\end{equation}

Thus if  $y=U(x)$ is such a function then $y=U(x)$ implies
$y=U(x+2\sqrt y)$ for all $(x,y)$ in some open domain $\Omega
\subseteq \mathbb R\times\mathbb R_+$. Let us start by considering
a more general situation when $y=U(x)$ is a branch of an algebraic
function, so that the graph $y=U(x)$ belongs to an irreducible
curve $F(x,y)=0$.  Making substitution $y=z^2$ we get a (possibly
reducible) curve $G(x,z)=F(x,z^2)=0$. Then the isochronicity means
that two complex algebraic curves $\Gamma: G(x,z)=0$ and $\Gamma':
G(x+2z,z)=0$ have a one-dimensional component in common. If
$\Gamma$ and, hence, $\Gamma'$ were irreducible, we would get
$\Gamma=\Gamma'$, so it would mean that for any $(x,z)\in\Gamma$,
$(x+2nz,z)\in\Gamma$ for any $n\ge 1$. This would mean that
$G(x,z)$ does not depend on $x$, which is not interesting.

Thus, $\Gamma$ must be reducible. Now, the only option is that
$G(x,z)=P(x,z)P(x,-z)$, where $P$ is irreducible polynomial.
Indeed since $G(x,z)$ is invariant under the involution $\tau:\
(x,z)\mapsto (x,-z)$, the irreducible factors must be interchanged
by $\tau$. We can have only one orbit, otherwise the product of
the factors along the orbits would give a factorization of $F$.

Since $G=G(x,z)$ and $G'=G(x+2z,z)$ must have a common divisor, we
have four cases to consider: either $P(x,z)=cP(x+2z,z)$ or
$P(x,z)=cP(x+2z,-z)$, or $P(x,-z)=cP(x+2z,z)$, or, finally,
$P(x,-z)=cP(x+2z,-z)$ for some constant $c$.

The first and the last cases are impossible unless $P$ does not
depend on $x$. In the second case, after passing to $u=x+z$ and
$v=z$, we get that the (still irreducible) polynomial
$Q(u,v):=P(u-v,v)$ has a symmetry $Q(u,v)=cQ(u,-v)$. Thus, either
$c=-1$ and $Q$ is odd in $v$, or $c=1$ and $Q$ is even in $v$. If
$Q$ is odd in $v$, it is reducible unless $Q=v$. In that case
$P=z$ which is not interesting. Thus, the only option is that
$c=1$ and $Q$ is any irreducible polynomial that is even in
$v$-variable. For example, taking $Q=v^2+u$ gives $P=z^2+x+z$,
$G(x,z)=(z^2+x+z)(z^2+x-z)$, and $F(x,y)=(y+x+\sqrt y)(y+x-\sqrt
y)=(y+x)^2-y=0$ defines an isochronous potential $y=u(x)$.

The harmonic oscillator corresponds to $Q=u$ which gives $P=x+z$,
$G(x,z)=(x+z)(x-z)$, so $F(x,y)=x^2-y$. Similarly, in the third
case, when $P(x,z)=cP(x-2z,-z)$, we pass to $u=x-z$ and $v=z$ to
get $Q(u,v):=P(u+v,v)=cP(u-v,-v)=cQ(u,-v)$, so as above we get
that $c=1$ and the irreducible polynomial $Q$ is even in
$v$-variable.

Now, returning to our original question when $U(x)$ is rational,
we have $F(x,y)=a(x)y+b(x)$ for some polynomials $a,b$. Then the
reducibility of $G(x,z)=az^2+b$ gives that the factors must be
linear in $z$, so up to a constant factor, $az^2+b=(\alpha+\beta
z)(\alpha -\beta z)$ where $\alpha, \beta$ are some polynomials in
$x$. As we saw above, we must have that either
$\alpha(x+2z)-\beta(x+2z)z\equiv \alpha(x)+\beta(x)z$ or
$\alpha(x-2z)-\beta(x-2z)z\equiv \alpha(x)+\beta(x)z$, identically
in $x,z$. Note that the second case reduces to the first by
changing $\beta$ to $-\beta$ and $z$ to $-z$. Thus, we have only
to consider the first functional equation:
\begin{equation*}
\alpha(x+2z)-\beta(x+2z)z\equiv \alpha(x)+\beta(x)z\,.
\end{equation*}
In that case, first differentiate with respect to $z$ at $z=0$, this gives
$\alpha'(x)=\beta(x)$. Next, put $x=0, z=t/2$, this gives:
$$\alpha(t)=\alpha(0)+t/2(\beta(t)+\beta(0))\,.$$ Now,
differentiating this with respect to $t$ and using the previous relation,
we get: $t\beta'(t)=\beta(t)-\beta(0)$, which gives that $\beta$
is linear polynomial, $\beta(x)=Cx+C_1$. By integrating, we find
that $\alpha(x)=1/2Cx^2+C_1x+C_0$. Now it is easy to check
directly that for such $\alpha,\beta$ the functional equation is
satisfied.

This gives us that $G(x,z)=\alpha^2-\beta^2z^2$ and
$u(x)=\alpha^2/\beta^2$. We have two cases: $C=0$, in which
$U(x)=x^2$ up to a shift in $x$, and the case $C\ne 0$, in which
we may assume that $C=1$ and find that up to a shift in $x$,
$U(x)=1/4x^2+g^2x^{-2}-g$ with $g=1/2C_1^2-C_0$. Note that $g=0$
is not allowed since in that case $P(x,z)$ and $F(x,y)$ are
reducible. This completes the proof of the Theorem.

\section{Discussion: quantum analogs.}

A natural quantum analog of our problem is to describe the
one-dimensional Schr\"odinger operators 

$$L = -\frac{d^2}{dx^2} + u(x)$$ with the equidistant
spectrum, i.e. isospectral to the usual harmonic oscillator. The
last problem was investigated by McKean and Trubowitz \cite{MT}
and Levitan \cite{L}. The relation between isospectrality and
isochronicity was discussed by Eleonski et al in \cite{EKK}. In
these papers the potential $U(x)$ was assumed to be regular on the
whole $x$-axis.

On the other hand there is "dressing chain" approach \cite{VS},
which allows to construct a wide class of the operators with the
spectrum consisting of several arithmetic progressions. The
corresponding potentials satisfy Painlev\'e-type equations and do
not belong to the functional class considered in \cite{MT,L} even
in the case when they are regular on the whole axis. But in
general the potentials produced by this approach have poles on the
real axis which makes the investigation of the spectral problem
less straightforward.

An interesting question therefore is to describe all singular rational
potentials with equidistant spectrum on one of the (possible two)
semi-lines. We conjecture that the answer is the same as in the
isochronicity problem, i.e. up to a shift such potentials have the
form $$u(x) = A x^2 + \frac {B} {x^2}.$$ 
The equidistance property of the corresponding spectra is well-known
in quantum mechanics (see e.g. Landau-Lifschitz \cite{LL2}, section 36).
An interesting novelty of the quantum case is that here parameter $B$ can be negative
(but larger than $- \frac{1}{4}$).

\section{Acknowledgements.} We are grateful to Holger Dullin and Sasha Pushnitski for useful  discussions.


\begin{thebibliography}{99}
\bibitem{A}
Appell P. {\it Traite de mechanique rationelle}, tome 1.
Gauthieres-Villars, Paris, 1902.
\bibitem{LL}
Landau L.D. and  Lifschitz  E.M. {\it Mechanics.} Pergamon, London,
1981.
\bibitem{U}
Urabe M. {\it Potential forces which yield periodic motions of a
fixed period.} J. Math. Mech.  {\bf 10}, 1961,  569-578.
\bibitem{CMV}
Cima A., Manosas F., and Villadelprat J. {\it Isochronicity for
several classes of Hamiltonian systems.} J. Diff. Eq. {\bf 157},
1999, 373-413.
\bibitem{C}
Calogero F.  {\it Solution of the one-dimensional N-body problems
with quadratic and/or inversely quadratic pair potentials.} J.
Math. Phys. {\bf 12}, 1971, 419-436.
\bibitem{OP}
Olshanetsky M.A., Perelomov A.M. {\it  Geodesic flows on symmetric
spaces of zero curvature, and explicit solutions of the
generalized Calogero model for the classical case. } Funkt. Anal.
Appl. {\bf 10}, 1976, no. 3, 86--87.
\bibitem{C2}
Calogero F. {\it Two new classes of isochronous Hamiltonian
systems.}  J. Nonlinear Math. Phys.  {\bf 11}  2004,  no. 2,
208--222.
\bibitem{MT}
McKean H. P. and Trubowitz E. {\it The spectral class of the
quantum-mechanical harmonic oscillator. } Comm. Math. Phys.  {\bf
82} , 1981/82, no. 4, 471--495.
\bibitem{L}
Levitan B. M. {\it Sturm-Liouville operators on the entire real
axis with the same  discrete spectrum. } Math. USSR-Sb.  {\bf 60},
1988,  no. 1, 77--106.
\bibitem{EKK}
Eleonski V.M., Korolev V.G. and Kulagin N.E. {\it On a classical
analog of the isospectral Schr\"odinger problem.} JETP Letters,
{\bf 65}, 1997, no.11, 889-893.
\bibitem{VS}
Veselov A.P. and Shabat A.B. {\it Dressing chain and spectral
theory of Schr\"odinger operators.} Funct. Anal. Appl., 27(2), 1993,
81-96.
\bibitem{LL2}
Landau L.D. and  Lifschitz  E.M.  {\it Quantum
Mechanics: Non-relativistic Theory.} Pergamon, Oxford, 1977.
\end{thebibliography}
\end{document}